\documentclass[journal]{IEEEtran}
\IEEEoverridecommandlockouts
\usepackage{cite}
\usepackage{amsmath,amssymb,amsfonts}
\usepackage{algorithmic}
\usepackage{graphicx}
\usepackage{textcomp,subcaption}
\usepackage{xcolor,multicol,multirow}
\usepackage{enumerate,textcomp,array,enumitem}
\usepackage{placeins}
\usepackage{threeparttable,url}
\usepackage{dblfloatfix}
\usepackage{flushend}
\usepackage{setspace}
\usepackage[running,displaymath]{lineno}

\newcommand*\linenomathpatch[1]{%
  \cspreto{#1}{\linenomath}%
  \cspreto{#1*}{\linenomath}%
  \csappto{end#1}{\endlinenomath}%
  \csappto{end#1*}{\endlinenomath}%
}

\linenomathpatch{equation}
\linenomathpatch{gather}
\linenomathpatch{multline}
\linenomathpatch{align}
\linenomathpatch{alignat}
\linenomathpatch{flalign}
\def\BibTeX{{\rm B\kern-.05em{\sc i\kern-.025em b}\kern-.08em
    T\kern-.1667em\lower.7ex\hbox{E}\kern-.125emX}}

\begin{document}

% \title{A Grid-Forming Inverter Active Power Exponential Droop Control for Improved Frequency Stability}

\title{Autonomous Grid-Forming Inverter Exponential Droop Control for Improved Frequency Stability}

\author{
R.~W.~Kenyon,~\IEEEmembership{Member,~IEEE}, 
A.~Sajadi,~\IEEEmembership{Senior~Member,~IEEE}, and
B.~M.~Hodge,~\IEEEmembership{Senior~Member,~IEEE}

\thanks{The authors are with the Renewable and Sustainable Energy Institute (RASEI) at the University of Colorado Boulder. 

Dr. Hodge and Dr. Kenyon are with the Electrical, Computer, and Energy Engineering department at the University of Colorado Boulder. 

Dr. Hodge and Dr. Sajadi are with the National Renewable Energy Laboratory in Golden, Colorado.}
}

% make the title area
\maketitle

\begin{abstract}
This paper introduces the novel \textit{Droop-e} grid-forming power electronic converter control strategy, which establishes a non-linear, active power--frequency droop relationship based on an exponential function of the power output. A primary advantage of \textit{Droop-e} is an increased utilization of available power headroom that directly mitigates system frequency excursions and reduces the rate of change of frequency. The motivation for \textit{Droop-e} as compared to a linear grid-forming control is first established, and then the full controller is described, including the mirrored inversion at the origin, the linearization at a parameterized limit, and the auxiliary autonomous power sharing controller. The analytic stability of the controller, including synchronization criteria and a small signal stability analysis, is assessed. Electromagnetic transient time domain simulations of the \textit{Droop-e} controller with full order power electronic converters and accompanying DC-side dynamics, connected in parallel with synchronous generators, are executed at a range of dispatches on a simple 3-bus system. Finally, IEEE 39-bus system simulations highlight the improved frequency stability of the system with multiple, \textit{Droop-e} controlled grid-forming inverters.
\end{abstract}


\section{Introduction}
\label{sec:introduction}

The large-scale integration of renewable energy, driven in part by world-wide carbon emission targets, lower marginal and capital costs, and environmental and air quality benefits, is primarily achieved with power electronic converters (PECs) that integrate variable renewable energy resources (such as wind and solar photovoltaics) into the AC power grid \cite{taylor_power_2016,kenyon_stability_2020}. These renewable driven PECs introduce new complexities into the operation and planning of power systems at all timescales \cite{milano_foundations_2018}, while also presenting the potential to engineer particular system-level dynamic responses that were historically driven in the short-term by the physics of synchronous generators (SGs). This work introduces the novel exponential droop control (\textit{Droop-e}) for grid-forming (GFM) PECs, leveraging their unique device level capabilities to improve the system level disturbance response of power systems of all sizes. 

The increased adoption of PECs in power systems has impacted the traditional stability regimes (voltage, rotor-angle, and frequency), while necessitating the introduction of two more (resonance and converter-driven) \cite{hatziargyriou_definition_2021}; this work focuses on frequency stability at the inertial and primary response time frames and the improvements possible with novel PEC controllers. It is well established that large penetrations of grid-following (GFL) PECs, the predominant contemporary type that explicitly require an exogenous reference phasor for operation, can lead to frequency instability \cite{ulbig_impact_2014}, while even grid-supporting mechanisms of these GFLs at high penetrations have been observed as insufficient to mitigate instabilities \cite{cheng_real-world_2023}. As a result, there has been substantial interest in the integration of GFM inverters that regulate the local voltage phasor and directly control the device frequency response \cite{north_american_electric_reliability_corporation_grid_2021}. Substantial research and experimentation has shown that these devices improve frequency stability in conjunction with SGs \cite{tayyebi_frequency_2020,ducoin_analytical_2024}, while also providing substantial benefit to the frequency response of power systems via increased damping \cite{sajadi_synchronization_2022}. However, at high shares of GFM devices with the most common linear droop control, the rate of change of frequency (ROCOF) is exacerbated and high frequency oscillations with other SGs can occur \cite{kenyon_interactive_2023}. 

% while commonly implemented active power limiting leads to large non-linearities in control response \cite{lasseter_grid-forming_2020}. 

A review of the state-of-the-art reveals that a variety of control schemes exist for the GFM approach \cite{rocabert_control_2012,du_comparative_2019, unruh_overview_2020}, including static (i.e., linear) droop \cite{piagi_autonomous_2006}, virtual synchronous machine \cite{beck_virtual_2007,poolla_placement_2019}, and virtual oscillator control \cite{johnson_synthesizing_2016}. Variations on linear droop include feed--forward mechanisms, virtual impedance loops for reactive power sharing, and adaptive droop with a power differentials \cite{tayab_review_2017}. The general concept of frequency shaping is presented in \cite{jiang_grid-forming_2021}. Matching control adjusts the device frequency based on the DC-link capacitor voltage state \cite{huang_virtual_2017}. Varied linear droop values are explored briefly in \cite{lasseter_grid-forming_2020}. The work in \cite{wang_exponential-function-based_2019} explored the concept of an exponential type droop control for reactive power sharing in resistive networks. Zhong \cite{zhong_robust_2013} investigates limitations in droop control based on resistive impedances. GFM power limiting with proportional-integral driven frequency reductions were presented in \cite{lasseter_grid-forming_2020} and \cite{pattabiraman_comparison_2018}. 

The novel, non-linear, power to frequency \textit{Droop-e} control provides multifaceted improvements including: 1) a larger utilization of available headroom, 2) a less deviant nadir and a more favorable rate of change of frequency (ROCOF), 3) improved frequency dynamics with a higher damping capability, and 4) a natural power limiting behavior. An auxiliary controller leverages the GFM direct frequency control to achieve autonomous power sharing amongst interconnected devices within the primary frequency response time frame. The \textit{Droop-e} control strategy represents immense potential for improving the frequency stability and resilience of emerging power grids by utilizing available energy resources more efficiently, especially battery energy storage systems (BESS) that are typically not operating near power limits \cite{conte_day-ahead_2020}.

An elementary version of the exponential control has been presented by the authors in \cite{kenyon_interactive_2023} and shown to improve the frequency stability of a Maui power system model following a generation loss in \cite{kenyon_using_2022}. This paper presents the fully developed \textit{Droop-e} controller accompanied by a full analytic stability assessment and a host of simulations depicting the range of possible operations. The remainder of the paper proceeds with a discussion on non-linear control motivation in Section \ref{sec:motivation}, a full controller description in Section \ref{sec:Droop-e}, analytical stability assessments in Section \ref{sec:analytic stability assessment}, and electromagnetic transient simulations on a simple 3-bus and the IEEE 39-bus system in Section \ref{sec: numerical simulations}. The presented control design, analysis, and simulation results firmly establish the capability of the \textit{Droop-e} control to improve the frequency stability of power systems at any penetration of PECs.

\section{Motivation for Non-Linear Frequency Control}
\label{sec:motivation}

This section summarizes the fundamental device frequency response of SGs and droop controlled GFMs that were rigorously analyzed in \cite{kenyon_interactive_2023}, highlighting the contrasts that serve as the motivational basis for the novel \textit{Droop-e} control.

\subsection{Linear Droop}

Following a transient load--generation imbalance, the SG first follows an inertial response, wherein the angular frequency ($d\delta_G/dt = \omega_G$) changes according to the swing equation \eqref{eq: frequency swing}:
\begin{align}
    \frac{2H}{\omega_{s}} \frac{d^2\delta_G}{dt^2} &= p_{m} - p_{e} - D(\frac{d\delta_G}{dt} - \omega_{s})\label{eq: frequency swing}
\end{align}
where $\delta_G$ is the generalized rotor angle, $H$ is the device inertia constant, $\omega_{s}$ is the synchronous frequency, $p_{m}$ is the rotational mechanical power, $p_{e}$ is the electrical power delivered to the network, and $D$ is the damping component. As $\omega_G$ changes, the governor and turbine adjust $p_{m}$ during the primary response, which is broadly described by \eqref{eq:SG governor}:
\begin{equation}
    T_{TG}\frac{dp_{m}}{dt} = - p_{m} + p_{set} - \frac{1}{M_D}\left(\frac{\omega_G}{\omega_{set}} - 1\right)\label{eq:SG governor}
\end{equation}
where $p_{set}$ is an exogenous active power setpoint, $M_D$ is the droop gain (i.e. 5\% in the United States), $\omega_{set}$ is the frequency setpoint, $T_{TG}$ is the turbine-governor time constant, and $\omega_G$ is the SG frequency. The inertial response period represents an uncontrolled exchange of rotational energy with the connected network. Comparatively, the controlled primary response of an SG is a reaction to a change in frequency. The amount by which $p_{m}$ changes is a control design; smaller $M_D$ values yield larger $p_{m}$ changes. Instability due to the increase in rate of change of $p_{m}$ occurs with small $M_D$ values \cite{kenyon_interactive_2023}. Setting $M_D=0$ is mathematically, and therefore physically, infeasible. 

The linear droop GFM frequency dynamics are shown in \eqref{eq: GFM freq} and \eqref{eq: GFM ROCOF}: 
\begin{align}
    \label{eq: GFM freq}\frac{d\delta_{I}}{dt} &=  M_D \left(p_{set} - p\right) + \omega_{set}\\
    \label{eq: GFM ROCOF}\frac{d\omega_I}{dt} &= -M_D\omega_{fil}\left(p - p_{meas}\right)
\end{align}
where $\delta_I$ is the inverter electric angle, $p_{set}$ is the exogenous power setpoint, $p$ is the filtered active power, $\omega_I$ is the inverter frequency, $\omega_{fil}$ is the power measurement cutoff frequency, and $p_{meas}$ is the measured, instantaneous power output. $\omega_I$ is not a control variable in the frequency dynamics of the GFM. Within the inertial and primary response periods, the frequency changes only after a change in power is registered; these are proactive devices. The value $M_D\omega_{fil}$ is congruent with the inverse of mechanical inertia ($1/H$); here it is considered an `effective inertia' as it represents the rate of change of frequency for this control type. Smaller $M_D$ values yield a greater effective inertia and therefore $M_D$ is a lever to influence both the change, and rate of change, of frequency.

\begin{figure*}
    \centering
    \includegraphics[width=1.6\columnwidth,trim={0 0 0 0},clip]{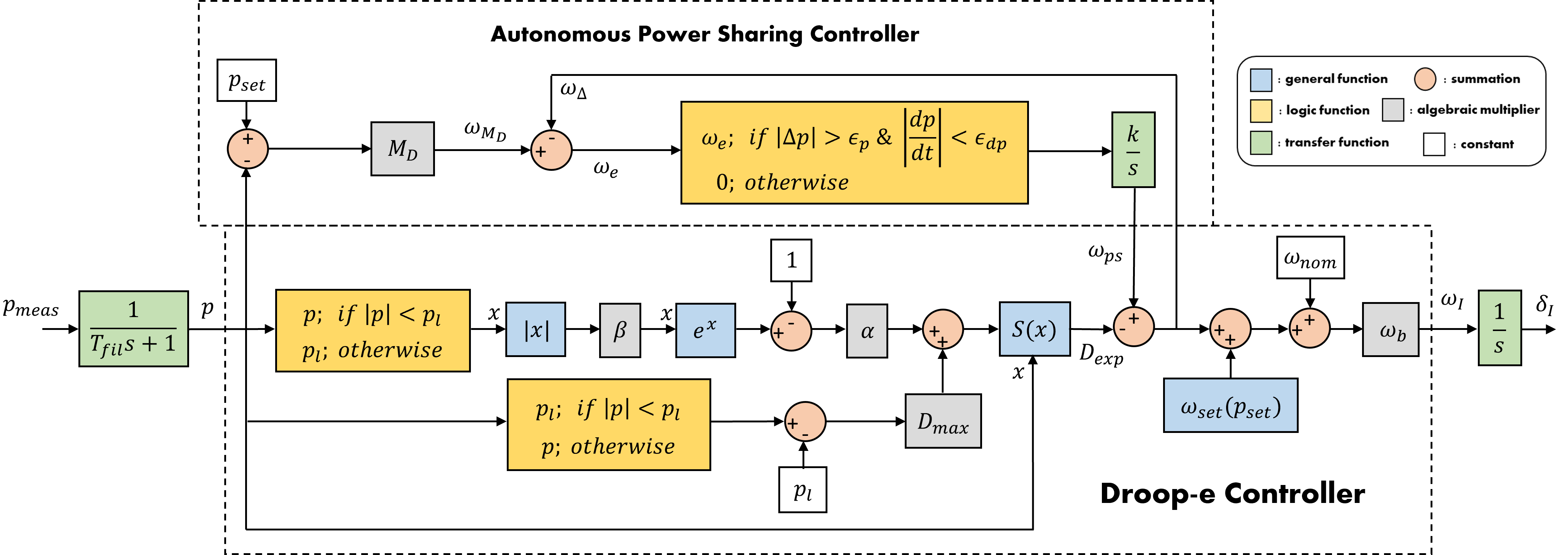}
    \caption{\small The full \textit{Droop-e} controller that receives the filter output power $p$, and calculates the output frequency $\omega_I$. The autonomous power sharing controller adds a frequency offset $\omega_{ps}$ to achieve power sharing within the primary frequency response period, only if this control objective is desired. The calculation of $p_l$ and $\omega_{set}$ are not explicitly shown.}
    \label{fig: Droop-e Controller}
\end{figure*}

\subsection{Concept of \textit{Droop-e}}

The primary idea behind the \textit{Droop-e} concept is supplanting the linear droop frequency control of a GFM device \eqref{eq: GFM freq} with an exponential function of power dispatch:
\begin{equation}
    \label{eq: GFM Droop-e freq}\frac{d\delta_{I}}{dt} = D_{e}(p,p_{set})
\end{equation}
where $D_e$ is the yet to be presented (in the preceding section) \textit{Droop-e} control. This approach allows a design wherein the change in frequency can be reduced for devices with large headroom, allowing them to exchange larger active power magnitudes with a network with corresponding smaller frequency deviations. This is accomplished by using $p$ as the independent variable in an exponential function. An exponential function is chosen instead of a power or quadratic because the tangent value does not go to zero when the argument is zero as with the others, a characteristic that has been shown to lead to instability in \cite{kenyon_interactive_2023}. An added benefit of the \textit{Droop-e} control is that the inertial response of the GFM, i.e., the resistance to a change in frequency with a change in power, is directly impacted by the tangent value of the exponential function at a particular operating point:
\begin{equation}
    \label{eq: GFM Droop-e rocof}\frac{d\omega_I}{dt} \propto \frac{\partial}{\partial p} D_e(p,p_{set})
\end{equation}
where $\propto$ is the proportional operator. The result is that because the \textit{Droop-e} control enables a larger use of headroom, this simultaneously improves the frequency stability of the system with a larger effective inertia. Additionally, near the power limits, the exponential curve is yielding an increased change in frequency with respect to power, incurring a natural device level limiting. The ability of the GFM device to deliver this active power rests on the assumption of an appropriately designed DC-link with the primary energy source and associated controller tuning. The work in \cite{kenyon_interactive_2023} has shown with rigorous simulation that with a sufficient design, this assumption of available active power is suitable for frequency control.

\section{The Droop-e Controller}
\label{sec:Droop-e}

This section establishes the core mathematics and rationale behind the \textit{Droop-e} controller, and accompanying secondary power sharing controller, with the block diagram shown in Fig. \ref{fig: Droop-e Controller}. The expression for \textit{Droop-e} control with the individual frequency components is Eq. \eqref{eq: Droop-e}:

\begin{equation}\label{eq: Droop-e}
    D_{e}(p,p_{set}) = \omega_b\left[\omega_{nom} + \omega_{set}(p_{set}) + D_{exp}(p) + \omega_{ps}\right]
\end{equation}
where $\omega_{nom}$ represents the nominal frequency setpoint, $\omega_{set}(p_{set})$ is the non-linear frequency offset, $D_{exp}(p)$ is the core exponential sliding mode frequency response, and $\omega_{ps}$ is the post transient autonomous power sharing component, all in per unit; $\omega_b$ is the base frequency of the system. The term $\omega_{set}(p_{set})$ is required to add an additional frequency offset due to the non-linear droop characteristics of the $D_{exp}(p)$ term. This section proceeds by addressing the exponential shaping in \ref{sec: shaping the exponential curve}, the inversion at the origin in \ref{sec: inverting the curve}, limiting the maximum droop value in \ref{sec: limiting the droop value}, and handling the frequency setpoint in \ref{sec: handling the setpoint}. Section \ref{sec: autonomous power sharing controller} presents the autonomous power sharing controller that adjusts the term $\omega_{ps}$ in Eq. \ref{eq: Droop-e} after a transient. The \textit{Droop-e} controller is parameterized with specific values in \ref{sec: full expression}, and accompanying curve traces are presented.

\subsection{Shaping the Exponential Curve}
\label{sec: shaping the exponential curve}

Considering first an active power argument domain of $p \in [0.0,1.0]$, the core exponential $E(p)$, as a function of $p$ is:
\begin{equation}
    \label{eq: exponential} E(p) =  -\alpha e^{\beta p}
\end{equation}
where $\alpha$ and $\beta$ are the linear and argument scalars, respectively. There is a negative sign explicitly in \eqref{eq: exponential} because the slope must always be negative for synchronization in an AC network \cite{jiang_grid-forming_2021}. $\alpha$ is selected to yield smaller droop values at lower dispatches, and $\beta$ is selected to drive a rapid increase in droop values near the dispatch limits. The first power derivative of $E(p)$ is:

\begin{equation}
    \label{eq: GFM rocof}\frac{d E(p)}{dp} = -\alpha\beta e^{\beta p}
\end{equation}
which expresses the `tangent droop value` at the operating point $p$. At low power dispatches, the $E(p)$ is primarily driven by the product $\alpha\beta$. To maximize headroom delivery at lower dispatches, while bounding the minimum droop value for stability, the following relationship is established:
\begin{equation}
    M_{min} \leq \alpha \beta < M_D
\end{equation}
where $D_{min}$ is a minimum droop value parameter. 

\subsection{Inverting the Curve}
\label{sec: inverting the curve}
To maintain the synchronization criteria, the curve must be mirrored and inverted at the origin ($p=0$). This inversion capability is constructed for an argument domain of $x\in [-1.0,1.0]$, which represents the active power output of a battery energy storage system. An analogous result is obtained for devices capable of only a positive power export ($p\in [0,1.0]$), with a projection $p_{control} = 2p - 1$, where $p_{control}$ is used in the \textit{Droop-e} formulation, and the inversion occurs at $p = 0.5$. All subsequent controller expressions are cast with $p$ only. The absolute value function ($|x|$) is defined as in \eqref{eq: absolute value}:
\begin{equation}\label{eq: absolute value}
    |x| := 
\begin{cases}
    x,& \text{if }  x\geq 0\\
    -x,  & \text{if } x < 0
\end{cases}
\end{equation}

The function $S(x)$ is defined as $S(x) := 2H(x) - 1$, where $H(x)$ is the Heaviside function ($H(x) = 1\phantom{0} \text{if }  x\geq 0, \phantom{0}=0\phantom{0} \text{otherwise}$). The output characteristic of $S(x)$ is \eqref{eq: sign function}:

\begin{equation}\label{eq: sign function}
    S(x) = 
\begin{cases}
    1,& \text{if }  x\geq 0\\
    -1,  & \text{if } x < 0
\end{cases}
\end{equation}

With the $|x|$ and $S(x)$ functions, the formulation in \eqref{eq: initial inversion} achieves an inverted reflection at the origin.

\begin{equation}\label{eq: initial inversion}
    E_{I}(p) = - S(p)\left[\alpha\left(e^{\beta|p|}-1\right)\right]
\end{equation}

\subsection{Limiting the Droop Value}
\label{sec: limiting the droop value}

Without limiting, the tangent droop value at maximum magnitude power outputs can be large and may be an undesirable operating characteristic. This is addressed by introducing a maximum droop value parameter, $D_{max}$. This is equal to the tangent droop value of $E_{I}$ at a particular operating power, defined as $p_{l}$. The power outputs at which $D_{max}$ occurs are at equal magnitude; i.e., $-p_l$ and $p_l$, due to the symmetry about the origin of \eqref{eq: initial inversion}. When operating past this $p_{l}$ value, the control reverts to linear droop as in \eqref{eq: GFM freq}, where $M_D = D_{max}$. The expression in \eqref{eq: exponential linearization} captures the inversion at the origin and the linearization for power outputs beyond $p_{l}$:

\begin{equation}\label{eq: exponential linearization}
D_{exp}(p) =
\begin{cases}
    - S(p)\left[\alpha\left(e^{\beta|p|}-1\right)\right],&\text{if }  |p| < p_l\\
    - S(p)\left[\alpha \left(e^{\beta p_l}-1\right) + D_{max} \Delta p\right],& \text{otherwise}
\end{cases}
\end{equation}
where $\Delta p = |p| - p_l$. The tangent droop value of $E_{I}$ is equal to $D_{max}$ when $|p| = p_l$. The derivative of $E_I$ with respect to $p$ is shown in \eqref{eq: Ei derivative}:
\begin{equation}\label{eq: Ei derivative}
    \frac{\partial E_{I}}{\partial p} = -\alpha\beta e^{\beta|p|}
\end{equation}
With \eqref{eq: Ei derivative} set equal to $D_{max}$, $p_{l}$ is calculated as:
\begin{equation}
    p_{l} = \frac{ln\left(\frac{D_{max}}{\alpha\beta}\right)}{\beta}
\end{equation}
with the negative sign dropped due to the origin symmetry.

\subsection{Handling the Frequency Setpoint}
\label{sec: handling the setpoint}

The non-linearity of $D_{exp}(p)$, and the linearization component, demand special attention to calculate the frequency setpoint ($\omega_{set}$) as a function of the power setpoint ($p_{set}$). Equation \eqref{eq: freq setpoint} shows the case specific per unit frequency calculation, that handles these complexities of $D_{exp}(p)$.

\begin{equation}\label{eq: freq setpoint}
    \begin{split}
    \omega&_{set}(p_{set}) = \\
    &\begin{cases}
        S(p_{set})\left[\alpha \left(e^{\beta|p_{set}|}-1\right)\right],&\text{if }  |p_{set}| < p_l\\
        S(p_{set})\left[\alpha\left(e^{\beta p_{l}}-1\right) + D_{max}\Delta p_{set}\right],& \text{otherwise}
    \end{cases}
    \end{split}
\end{equation}
where $\Delta p_{set} = |p_{set}| - p_l$. For $p_{set}$ values with a magnitude less than $p_l$, the frequency value along the exponential curve is calculated and added as an offset for the frequency setpoint. For $p_{set}$ values larger than $p_l$, the linearization handling is captured in a similar manner. Note that the frequency setpoint is only necessary to calculate the correct setpoint at a specific power setpoint, it has no impact on the dynamic response of the \textit{Droop-e} control.

\subsection{Autonomous Power Sharing Controller}
\label{sec: autonomous power sharing controller}
By design, the \textit{Droop-e} control deviates from transient power sharing in the inertial and primary frequency response periods. The autonomous power sharing controller, shown in Fig. \ref{fig: Droop-e Controller}, operates by adding the offset component $\omega_{ps}$ to achieve equitable power sharing after the transient. With an offset instead of a full-bypass, the \textit{Droop-e} control continues to provide damping to the system with $D_{exp}(p)$, but the added offset $\omega_{ps}$ changes $\omega_I$ and causes frequency responsive devices on the network to react. This power sharing controller is presented as a method to achieve autonomous power sharing within the primary frequency response period. However, this controller can be easily bypassed if automatic governor control methods are preferred in the secondary response period.

The frequency deviation that would result with a linear droop is calculated with the relation \eqref{eq: GFM freq}, ignoring the frequency setpoint \eqref{eq: static droop ps}:
\begin{equation}\label{eq: static droop ps}
    \omega_{M_D} = (p_{set}-p)M_D
\end{equation}
This frequency is compared with $D_{exp}$ and the current power sharing component $\omega_{ps}$ to generate an error, $\omega_e$:
\begin{equation}\label{eq: freq error}
    \omega_{e} = \omega_{M_D} + D_{exp} - \omega_{ps}
\end{equation}

This error is passed through a logic block only after a disturbance is registered, and the transients have diminished:
\begin{equation}\label{eq: disturbance criteria}
    \omega_{e,out} = 
    \begin{cases}
        \omega_e ,&\text{if } |\Delta p|>\epsilon_p \land |\frac{dp}{dt}| < \epsilon_{dp}\\
        0,& \text{otherwise}
    \end{cases}
\end{equation}
where are $\epsilon_p$ and $\epsilon_{dp}$ are disturbance tolerance parameters. Once the disturbance criteria are met, this error is passed through an integrator block with gain $k$, which generates the frequency offset, $\omega_{ps}$. As this offset is added to the output frequency $\omega_I$, the GFM power output will change due to the dynamics of AC power transfer, causing following changes in $\omega_{D_e}$. The GFM will arrive at the power sharing value as specified by $M_D$ as $\omega_e$ is driven to $0$ by the integrator.

\begin{table}[htb]
    \footnotesize
    \centering
    \caption{\small Selected \textit{Droop-e} Parameters}
    \label{tab: Droop-e parameters}
    \begin{tabular}{c||c}
        Parameter & Value \\\hline\hline
        $\alpha$ & 0.0012 \\
        $\beta$ & 3.2\\
        $D_{min}$ & 0.25\%\\
        $D_{max}$ & 6.0\%\\
        $\omega_b$ & 376.98 radians (60 Hz)\\
        $p_l$ & 0.86 pu \\
        $min\phantom{0}D_e$ = $\alpha\beta$ & 0.45\%\\
        $M_D$ & 5\%\\
        $k$ & 0.2\\
        $\epsilon_p$ & 0.01\\
        $\epsilon_{dp}$ & 0.001
    \end{tabular}
\end{table}

\begin{figure}[h]
    \centering
    \includegraphics[width=1.0\columnwidth,trim={0 0 0 40},clip]{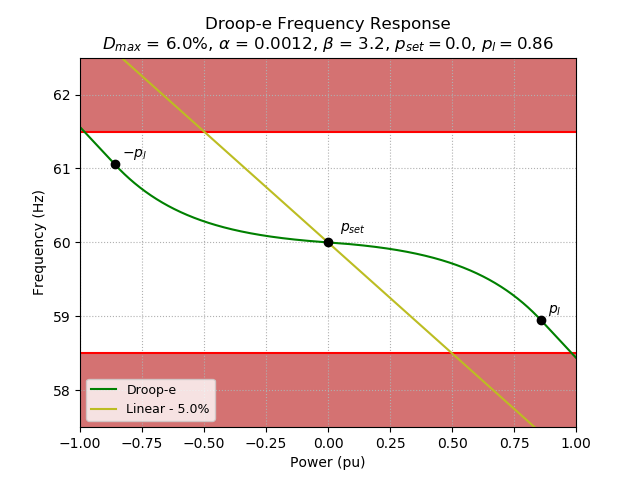}
    \caption{\small \textit{Droop-e} frequency curve  as a function of power output for the parameters specified in Table \ref{tab: Droop-e parameters}, compared to a 5\% linear droop.}
    \label{fig: Droop-e}
\end{figure}

\subsection{Full \textit{Droop-e} Expression}
\label{sec: full expression}

Table \ref{tab: Droop-e parameters} shows the \textit{Droop-e} parameter values used for this study, which were selected to achieve a full active power output range from -1.0 to 1.0 within a frequency range of 61.5 to 58.5 Hz. Here, a $D_{max}$ of 6\% was selected to allow a natural power limiting near the peak outputs. The power--frequency curve of the \textit{Droop-e} control with these parameters at a power dispatch of $p_{set}=0.0$ is shown in Fig. \ref{fig: Droop-e}. It is evident that the Droop-e curve has the mirrored inversion at the origin, and the linearization when $|p|>p_l$.

\begin{figure}[h]
    \centering
    \includegraphics[width=0.8\columnwidth,trim={0 0 0 40},clip]{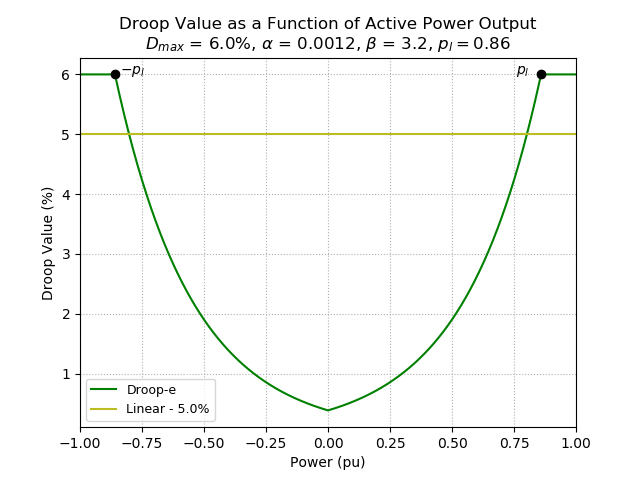}
    \caption{\small\textit{Droop-e} tangent droop value across the full range of power output $p$ with parameters from Table \ref{tab: Droop-e parameters}. Note the limits for $|p|>p_l$ and the continuous value across the range of outputs.}
    \label{fig: Droop-off}
\end{figure}

Figure \ref{fig: Droop-off} shows the tangent droop value as a function $p$ of the \textit{Droop-e} curve from Fig. \ref{fig: Droop-e}. This tangent droop is inversely proportional to the effective inertia per \eqref{eq: GFM rocof}. Clearly, the \textit{Droop-e} control provides a larger effective inertia to the system for the majority of dispatch points as compared to the linear 5\% droop. As indicated, the droop value is limited to 6\% through the linearization of the controller at values greater than $\pm p_l$.

\section{Analytical Stability Assessment of \textit{Droop-e}}
\label{sec:analytic stability assessment}

This section presents the analytical stability analyses of the \textit{Droop-e} controller. 

\subsection{Synchronization Criteria}

The power derivative of the \textit{Droop-e} controller $D_{e}$ across the full power output domain $p\in [-1.0,1.0]$ is:
\begin{equation}\label{eq: Dexp derivative}
\frac{\partial D_{exp}}{\partial p} =
\begin{cases}
    -\alpha\beta e^{\beta|p|},&\text{if} \phantom{0} |p| < p_l\\
    - D_{max},& \text{otherwise}
\end{cases}
\end{equation}

The derivative of $D_{exp}$ exists at all points on the power output domain. Furthermore, because $p_l$ is strictly defined as the power output where \eqref{eq: Ei derivative} is equal to $D_{max}$, the derivative is continuous at all points across the output domain. Therefore, \textit{Droop-e} is continuous and smooth across the entire output domain, indicating that there are no discontinuities in the control response. The expression in \eqref{eq: Dexp derivative} proves that the droop value is negative across the entire power output domain. This is a necessary criteria for synchronization in AC networks that have intrinsic, natural droop characteristics \cite{sajadi_synchronization_2022}; i.e., an increase/decrease in power output causes a decrease/increase in device frequency. Therefore, across the entire power output domain, the frequency is monotonically decreasing and has guaranteed stability in inductive AC networks \cite{jiang_grid-forming_2021}.

\subsection{Stability of the Power Sharing Controller}

The autonomous power sharing controller is a closed-loop feedback control system with the control objective to add a frequency offset ($\omega_{ps}$) to $\omega_I$ to achieve a parameterized power sharing droop value, $M_D$. This process achieves equitable power sharing among all frequency response-supporting generators in the system, and is activated only after primary frequency oscillations are dissipated and its value has reached an equilibrium. The transfer function that describes this control process is given by:
\begin{equation}\label{eq: power sharing controller laplace}
    \Delta \omega (s)=\frac{D_{exp}(p) s+k M_D\Delta p}{s+k}
\end{equation}
where $\Delta \omega=\omega_I-\omega_{set}$ and $\Delta p=p_{set}-p$. The limit of \eqref{eq: power sharing controller laplace} is considered for two cases. First, the limit as $s\rightarrow\infty$, which is the case for GFM power output transients \eqref{eq: power control limit inf}:

\begin{equation}\label{eq: power control limit inf}
    \lim_{s\to \infty} \Delta\omega(s) = D_{exp}(p)
\end{equation}
This limit shows that during transients, which may occur after the autonomous power sharing controller is initiated, core exponential relation remains dominate. The limit of $s\rightarrow 0$ is shown in \eqref{eq: power control limit 0}:

\begin{equation}\label{eq: power control limit 0}
    \lim_{s\to \infty} \Delta\omega(s) = M_D\Delta p
\end{equation}
which indicates that the integrator in the autonomous power sharing controller will enforce the added frequency offset and achieve a change in frequency commensurate with the desired, equitable droop value, $M_D$.

\subsection{Small Signal Stability Analysis}
\label{sec:SSSA}

Small signal stability analysis (SSSA) is a linearization technique that is applied to a differential--algebraic equation set of a system with $m$ differential states ($x_m$) and $n$ algebraic variables ($y_n$) \cite{sauer_power_2017}. This system is linearized in the form of:
\begin{equation}\label{eq: linearized system}
    \Delta \dot{x}_m = A_{sys}\Delta x_m
\end{equation}
where $A_{sys}$ is the system state matrix that contains the algebraic variables. The complex eigenvalues of $A_{sys}$ are $\lambda_m = \alpha_m + j\omega_m$, which capture the time-evolution of the system \eqref{eq: linearized system}. Positive real parts ($\alpha_i >0$) represent fundamental instability. The damping component ($\zeta_m$) of each eigenvalue is calculated as $\label{eq: damping}\zeta_m = \frac{-\alpha_m}{\sqrt{\alpha_m^2 + \omega_m^2}}$.

\begin{figure*}[htbp]	 
	\centering
	\begin{subfigure}[t]{2.13in}
		\centering
		\captionsetup{justification=centering}
		\includegraphics[trim=3 4 110 40, clip,width=1\textwidth]{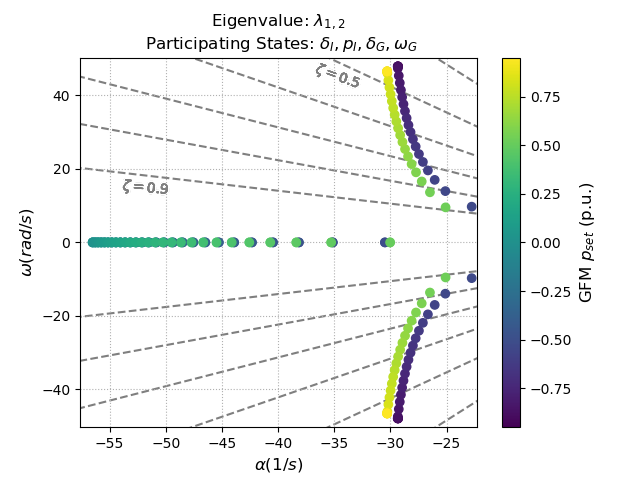}
	    \caption{\small Eigenvalue: $\lambda_{1,2}$\\ States: $\delta_I$, $p_{m,I}$, $\delta_G$, and $\omega_G$}\label{fig: eigen 1}
    \end{subfigure}
	\hfill
	\begin{subfigure}[t]{2.13in}
		\centering
		\captionsetup{justification=centering}
		\includegraphics[trim=3 4 110 40,clip, width=1\textwidth]{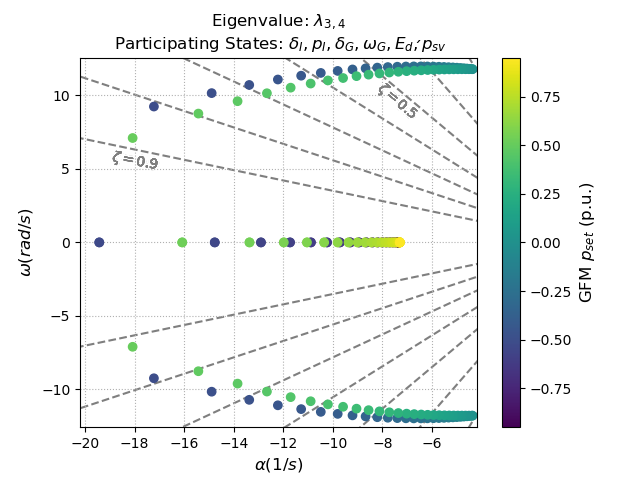}
		\caption{\small Eigenvalue: $\lambda_{3,4}$\\ States: $\delta_I$, $p_{m,I}$, $\delta_G$, $\omega_G$, $E'_d$, and $p_{SV}$.}\label{fig: eigen 2}
	\end{subfigure}
	\hfill
	\begin{subfigure}[t]{2.62in}
	    \centering
	    \captionsetup{justification=centering}
	    \includegraphics[trim=3 4 30 40,clip, width=1\textwidth]{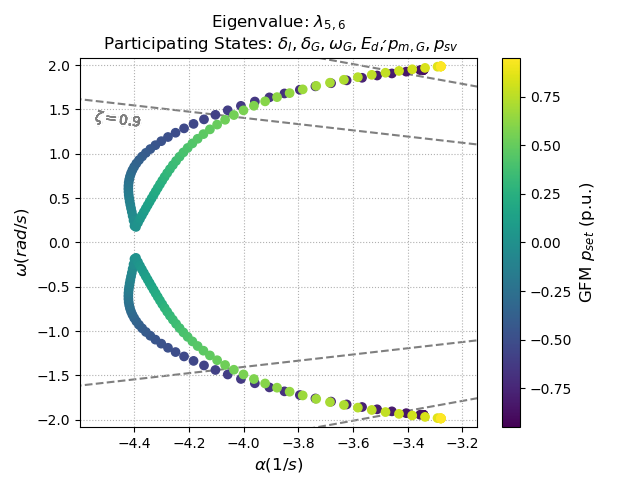}
	    \caption{\small Eigenvalue: $\lambda_{5,6}$\\ States: $\delta_I$, $\delta_G$, $\omega_G$, $E'_d$, $p_{m,G}$, and $p_{SV}$}\label{fig: eigen 3}
	\end{subfigure}
	\caption{\small Eigenvalue trajectories of the simple 3-bus system for those with GFM state participation. Note the varied x and y axis scales.}\label{fig: eigenvalues}
\end{figure*}

\begin{equation}
    \begin{split}
        \label{eq: states}
        x_{SG} &= [\delta_{G}, \omega_{G}, E'_q, E'_d, E_{fd}, V_R, R_f, p_{m,G}, p_{SV}]^T\\
        x_{GFM} &= [\delta_{I}, p_I]^T
    \end{split}
\end{equation}

\begin{figure}[ht]
    \centering
    \includegraphics[width=0.6\columnwidth,trim={0 0 0 0},clip]{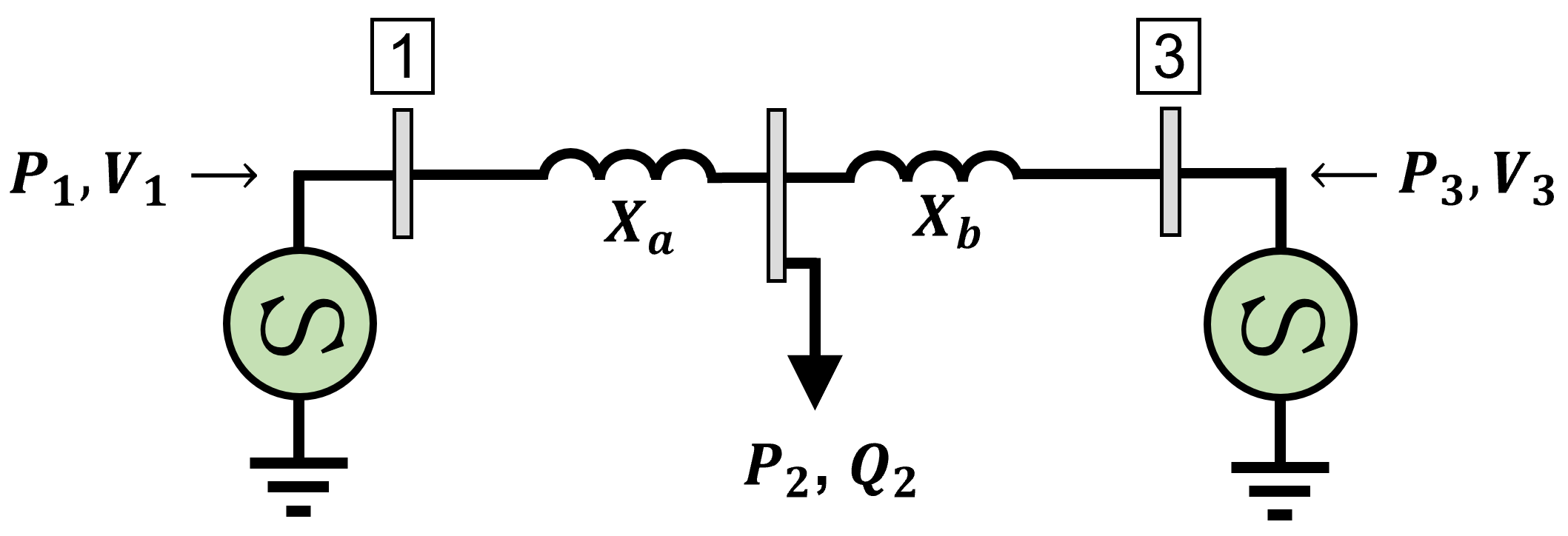}
    \caption{\small The simple 3-bus system with a synchronous generator located at bus 1 and a \textit{Droop-e} grid-forming PEC at bus 3.}
    \label{fig: three bus model}
\end{figure}

\begin{figure}[ht]
    \centering
    \includegraphics[width=0.8\columnwidth,trim={0 0 0 0},clip]{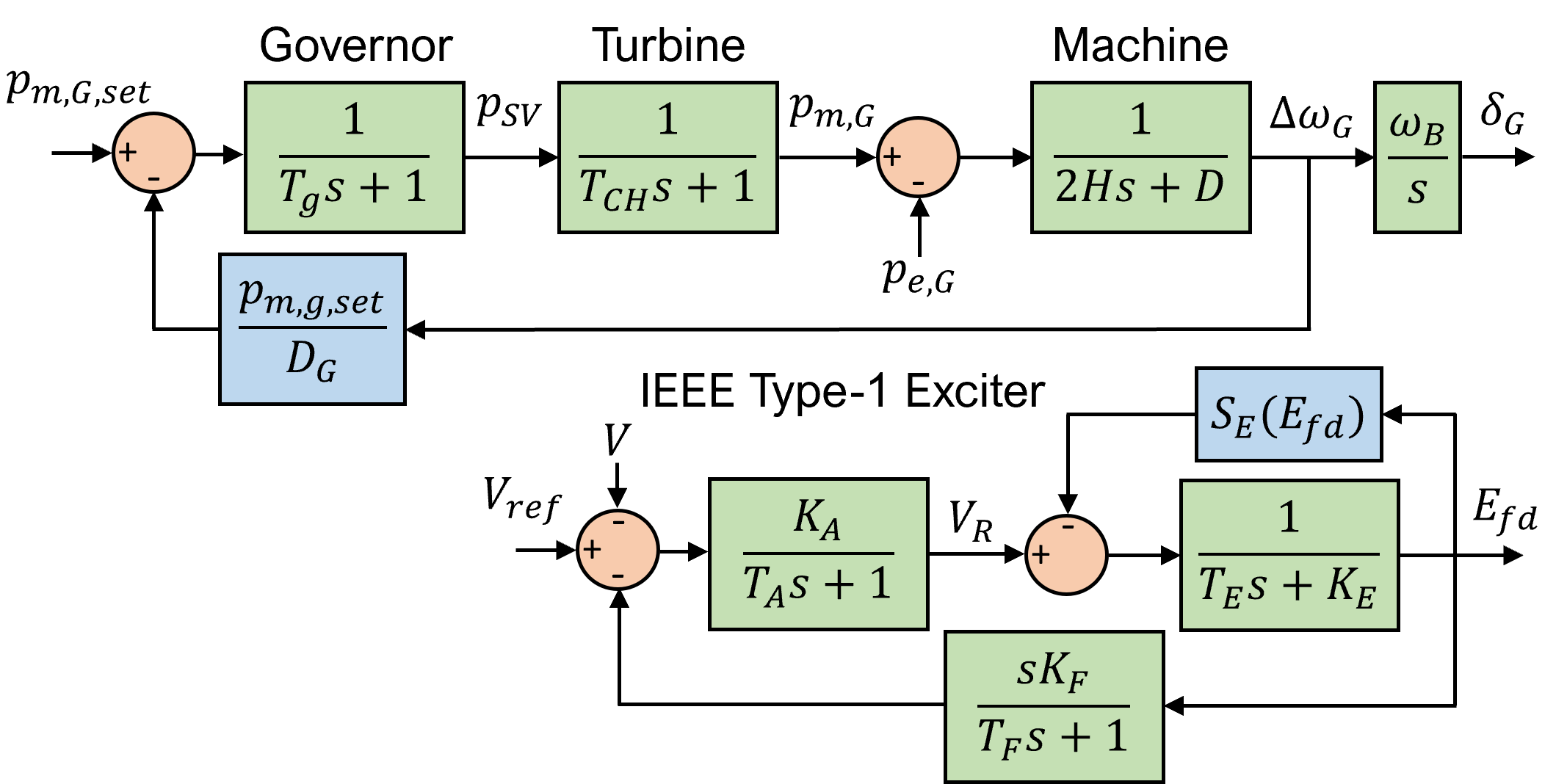}
    \caption{\small The synchronous generator model with governor, turbine, machine, and exciter dynamic sub-systems.}
    \label{fig: SG Block}
\end{figure}

A simple 3-bus network (shown in Fig. \ref{fig: three bus model}) is used for this SSSA (and subsequently in the simulations). A 100-MVA SG is located at bus 1. A battery backed ($p\in [-1.0,1.0]$), \textit{Droop-e} controlled, 50-MVA GFM PEC is at bus 3. The reactances $X_a = X_b = 0.05 pu$ represent the interconnecting lines. The network bases are $S_b = 100 MVA$ and $V_b = 18 kV$. The load at bus 2 is constant power ($P_2 = 75 MW$, $Q_2 = 25 Mvar$). The SG model used is the base model from \cite{sauer_power_2017}, represented by the block diagram of Fig. \ref{fig: SG Block}. The saturation function is an exponential of the form: $S_E(E_{fd}) = \gamma e^{\epsilon E_{fd}}$. Flux decay is included. The states of the 9th order model are \eqref{eq: states}. The parameters are from the 9-bus model SG 3 in \cite{sauer_power_2017}, with governor and turbine parameters from \cite{pattabiraman_comparison_2018}. 

The dynamic model of the GFM is as shown in Fig. \ref{fig: Droop-e Controller}. This model has two states, shown in \eqref{eq: states}, where the subscript $I$ has been added. A voltage behind impedance model, with the output impedance of the LCL filter, is the circuit coupling the dynamic model to the network \cite{pattabiraman_comparison_2018}. The internal voltage and current controllers are not modeled, and a constant voltage at the LCL capacitor is assumed \cite{kenyon_open-source_2021}. It is recognized that the internal controllers can impact small signal stability \cite{markovic_understanding_2021,xiong_modeling_2020}, but these oscillatory modes are outside of the scope of this study, which is focused on the slower mode SG interactions; regardless, the accompanying electromagnetic transient simulations were executed with full order GFM models. The linearization process is executed as in \cite{sauer_power_2017}. All network details are provided in Table \ref{tab: system parameters}.

\subsection{SSSA Results}

The eigenvalues of the 3-bus system of Fig. \ref{fig: three bus model} were calculated for at the full per unit dispatch range of the battery backed, \textit{Droop-e} controlled GFM inverter. Three complex eigenvalue pairs were identified via participation factor analysis as involving the GFM states of; $\lambda_{1,2}$, $\lambda_{3,4}$, and $\lambda_{5,6}$. The eigenvalue pair $\lambda_{1,2}$ involves the states $\delta_I$, $p_{m,I}$, $\delta_G$, and $\omega_G$. Figure \ref{fig: eigen 1} shows that the eigenvalue trajectory is pure real and negative, for the range $p_{set} = [-0.4, 0.4]$, and at $p_{set} = \pm 0.4$ a bifurcation occurs where modes transition from pure real to oscillatory. The damping decreases monotonically up to the full dispatch point, $p_{set} = \pm 1.0$, below $\zeta = 0.45$. Figure \ref{fig: eigen 2} shows the evolution of eigenvalue pair $\lambda_{3,4}$ in the left half plane, which includes participation from the states $\delta_I$, $p_{m,I}$, $\delta_G$, $\omega_G$, $E'_d$, and $p_{SV}$. This eigenvalue pair has an oscillatory element at low $p_{set}$ values that becomes increasingly damped as the dispatch magnitude increases. The mode becomes pure real at the same bifurcation point as $\lambda_{1,2}$, indicating a coupling between $\lambda_{1,2}$ and $\lambda_{3,4}$. As both modes are higher frequency and contain participation from $p_{m,I}$, but not the SG mechanical power ($p_{m,G}$), these eigenvalues driven primarily by the first-order low-pass power filter measurement of the GFM device.

The final eigenvalue pair with GFM state participation is $\lambda_{5,6}$ shown in Fig. \ref{fig: eigen 3}. This eigenvalue pair includes both the GFM and SG angles, and the $p_{m,G}$, and it is a low frequency mode at $0.06\to0.63$ Hz. The trajectory of this pair corroborates the frequency damping benefit to the system of the \textit{Droop-e} control, particularly at low power dispatches \cite{kenyon_interactive_2023}. This mode depicts a strongly damped oscillatory mode at all dispatches, with a decreasing damping as the dispatch of the GFM inverter is increased. All three eigenvalues with participation from the \textit{Droop-e} controlled GFM device display appropriate damping, indicating that the controller is stable when connected in parallel with a SG device across the range of possible dispatches.

\begin{figure*}[htbp]	
	\centering
	\begin{subfigure}[t]{2.35in}
		\centering
		\includegraphics[trim=3 6 38 41,clip,width=1\textwidth]{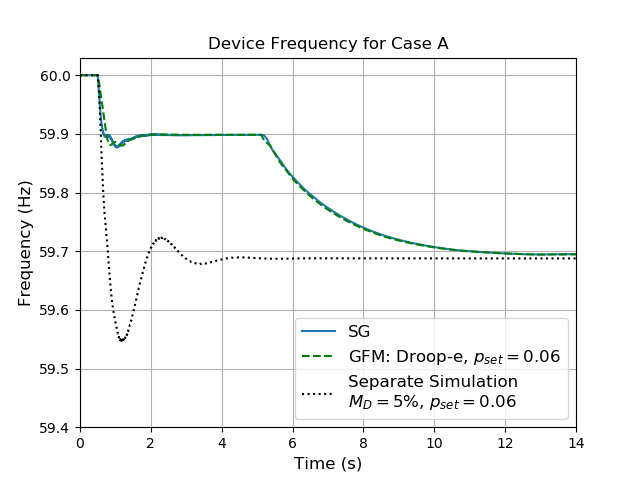}
		\caption{\small Case A - Frequency}\label{fig: freq a}		
	\end{subfigure}
        \hfill
	\begin{subfigure}[t]{2.35in}
		\centering
		\includegraphics[trim=3 6 38 37,clip, width=1\textwidth]{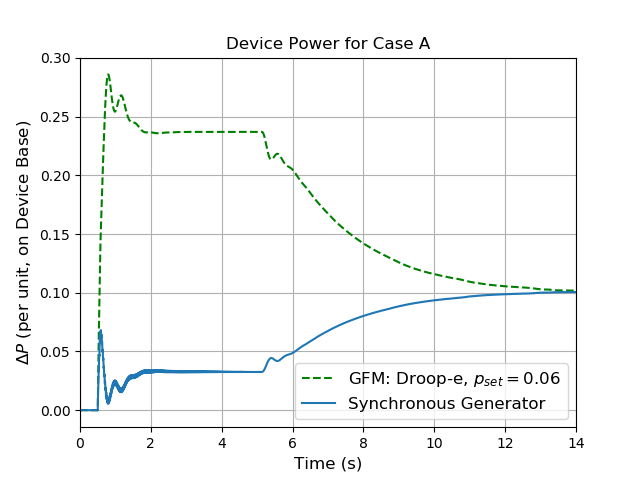}
		\caption{\small Case A - Power}\label{fig: power a}
	\end{subfigure}
	\hfill
	\begin{subfigure}[t]{2.35in}
	    \centering
	    \includegraphics[trim=3 6 38 37,clip, width=1\textwidth]{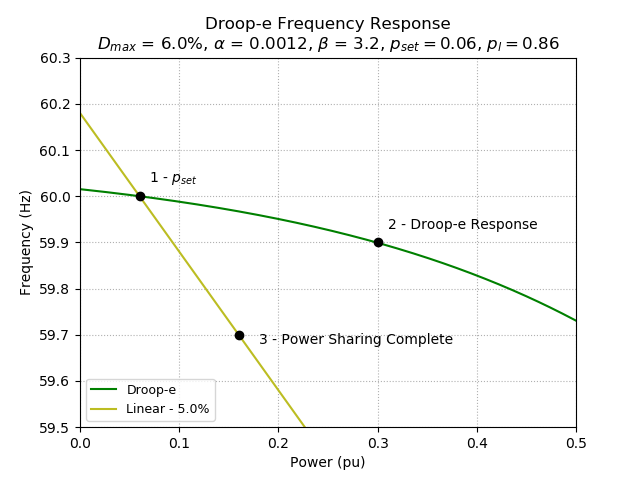}
	    \caption{\small Case A - GFM Droop Trajectory}\label{fig: droop a}
	\end{subfigure}
	\caption{The frequency response, power response, and droop trajectory for the GFM and SG devices in Case A.}\label{fig: Case A 3bus} 
\end{figure*}

\section{Electromagnetic Transient Simulations}
\label{sec:simulations}

This section details the electromagnetic transient dynamic simulation experiments conducted on the simple 3-bus, and the IEEE 39-bus test systems to assess the time-domain response of the \textit{Droop-e} control in a variety of operating conditions. These simulations were conducted in the PSCAD software package \cite{manitoba_hydro_international_ltd_power_nodate}. The GFM PEC is represented with a core 13th-order model including internal current and voltage controllers, an output LCL filter, and power filtering. A 4th order battery energy storage system model is accompanied by a 2nd order buck-boost controller to regulate the DC-link voltage of the core PEC model. The GFM device is executing a reactive power-voltage droop control ($Q-V$). A full description of the models and networks can be found in references \cite{kenyon_open-source_2021} and \cite{kenyon_interactive_2023}. The SG model was constructed with internal PSCAD models. The device network parameters are in Appendix \ref{sec:appendix}, Table \ref{tab: system parameters}. All models and networks are available open-source at \cite{kenyon_pypscad_2020}. In all cases, the simulations are executed by first bringing the system into steady state prior to releasing all dynamic elements and applying the perturbation.

The frequency nadir, and peak ROCOF, which is calculated with a sliding window of $T_w = 0.1s$;  $\label{eq:ROCOF} max|\dot{f_G}(t)| = \frac{\omega_G(t + T_{w}) - \omega_G(t)}{2\pi T_{w}}$, are used to characterize the system response. The frequency statistics are calculated from the SG rotational frequency alone for the 3-bus results. On the 39-bus system, the weighted average frequency is used for statistics - $\label{eq:measurefrequency}
    f(t) = \frac{\sum_{i=1}^n (MVA_i*f_i(t))}{\sum_{i=1}^n MVA_i}
$ where $f_i(t)$ is the frequency of device $i$ at time $t$, $MVA_i$ is the device $i$ rating, and $n$ is the number of devices. The mechanical inertia rating of the system configuration, presented in Table \ref{tab: 39 cases results}, is a weighted average calculated as $\label{eq:aggregate inertia} H = \frac{\sum_{i=1}^n H_i S_{B,i}}{\sum_{i=1}^nS_{B,i}}$
where $H_i$ is the inertia rating (in $s$) of device $i$, $S_{B,i}$ is the MVA rating of device $i$, and $n$ is the number of devices. $H_i = 0$ for GFM devices. The primary mode frequency and damping were calculated with the matrix pencil method \cite{sarkar_using_1995}.

\subsection{3-Bus System}
\label{sec:3 bus simulation}

The 3-bus system topology is identical to Fig. \ref{fig: three bus model}. Three simulations with $\pm$20\% load perturbations were executed to capture the full diversity of the \textit{Droop-e} control. Case `A' captures a load increase case for a near-zero dispatch of the GFM and case `B' a load increase for a highly dispatched GFM. Case `C' is a load decrease for a positive, near-zero dispatch, capturing the inversion of the \textit{Droop-e} curve through the origin. The dispatch conditions and applied load step are summarized in Table \ref{tab: 3 bus cases}.

\begin{table}[htb]
    \footnotesize
    \renewcommand{\arraystretch}{1.2}
    \setlength{\tabcolsep}{.3em}
    \centering
    \caption{\small Dispatches for each case on a 100 MVA base.}
    \label{tab: 3 bus cases}
    \begin{tabular}{c||c|c||c|c||c}
         \multirow{2}{*}{\rotatebox{0}{Case}} & \multicolumn{2}{c||}{SG} & \multicolumn{2}{c||}{Droop-e GFM} & Load Step\\
         & \phantom{0}P (pu)\phantom{0} & Q (pu) & \phantom{0}P (pu)\phantom{0} & Q (pu) & $\Delta S$\\
        \hline\hline
        A & 0.73 & 0.21 & 0.03 & 0.07 & 20\%\\
        B & 0.40 & 0.24 & 0.40 & 0.03 & 20\%\\
        C & 0.73 & 0.21 & 0.03 & 0.07 & -20\%\\
    \end{tabular}
\end{table}

\begin{table}[tb]
    \footnotesize
    \renewcommand{\arraystretch}{1.0}
    \setlength{\tabcolsep}{.3em}
    \centering
    \caption{\small Case Results--All apply prior to power sharing ($\approx$ 5s); frequency statistics apply to synchronous generator shaft speed; $\Delta P$ is on 100 MVA base.}
    \label{tab: cases results}
    \begin{tabular}{c||c|c||c|c|c}
        \multirow{2}{*}{\rotatebox{0}{Case}} & $\Delta P_{SG}$ & $\Delta P_{GFM}$ & Nadir & ROCOF & Damping \\
         & (pu) & (pu) & (Hz) & (Hz/s) & ($\zeta$)\\
        \hline\hline
        A & 0.033 & 0.119 & 59.9 & 0.77 & 0.47 \\
        B & 0.106 & 0.045 & 59.52 & 1.48 & 0.36 \\
        C & -0.024 & -0.127 & 60.09 (Over Frequency) & 0.68 & 0.52 \\
    \end{tabular}
\end{table}

\begin{figure*}[htbp]	
	\centering
	\begin{subfigure}[t]{2.35in}
		\centering
		\includegraphics[trim=3 6 38 41,clip,width=1\textwidth]{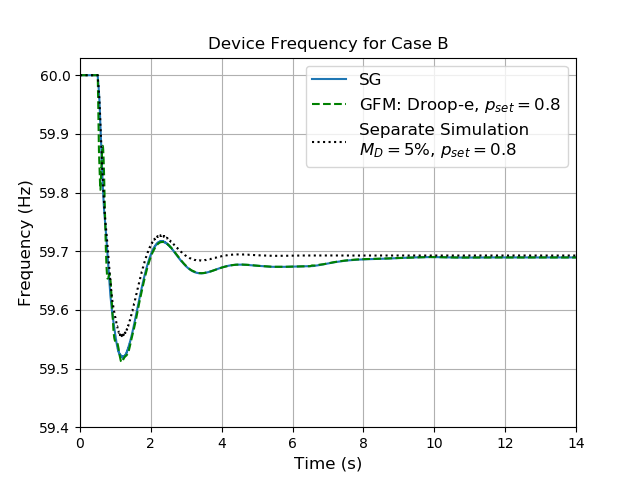}
		\caption{\small Case B - Frequency}\label{fig: freq b}		
	\end{subfigure}
    \hfill
	\begin{subfigure}[t]{2.35in}
		\centering
		\includegraphics[trim=3 6 38 41,clip, width=1\textwidth]{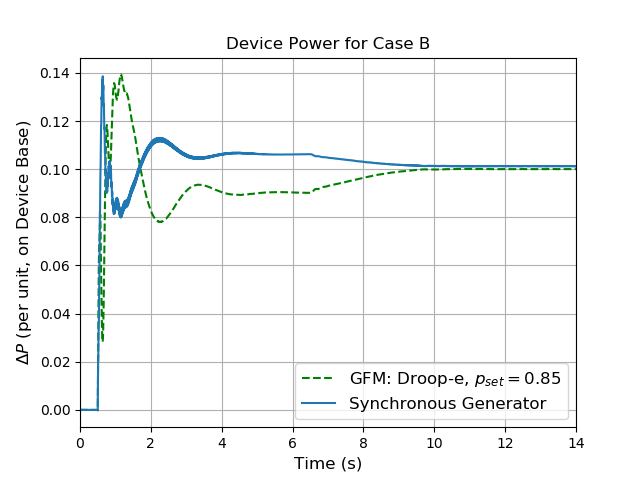}
		\caption{\small Case B - Power}\label{fig: power b}
	\end{subfigure}
	\hfill
	\begin{subfigure}[t]{2.35in}
	    \centering
	    \includegraphics[trim=3 6 38 37,clip, width=1\textwidth]{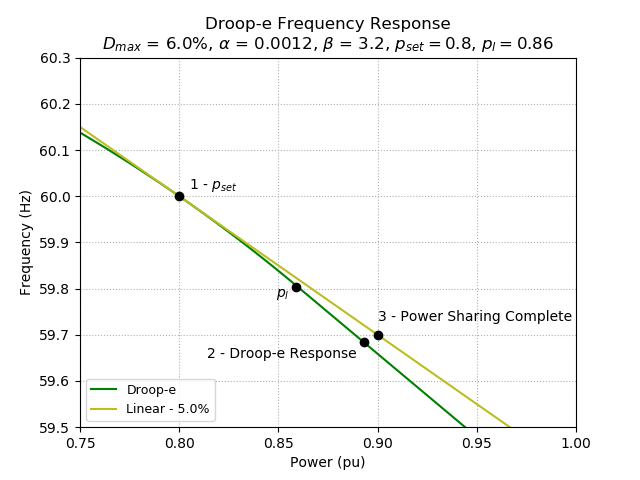}
	    \caption{\small Case B - GFM Droop Trajectory}\label{fig: droop b}
	\end{subfigure}
	\caption{The frequency response, power response, and droop trajectory for the GFM and SG devices in Case B.}\label{fig: Case B 3bus} 
\end{figure*}

\begin{figure*}[htbp]	
	\centering
	\begin{subfigure}[t]{2.35in}
		\centering
		\includegraphics[trim=3 6 38 41,clip,width=1\textwidth]{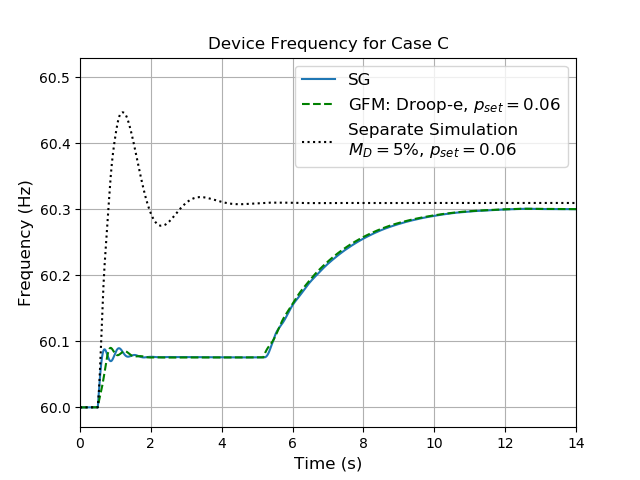}
		\caption{\small Case C}\label{fig: freq c}		
	\end{subfigure}
    \hfill
	\begin{subfigure}[t]{2.35in}
		\centering
		\includegraphics[trim=3 6 38 41,clip, width=1\textwidth]{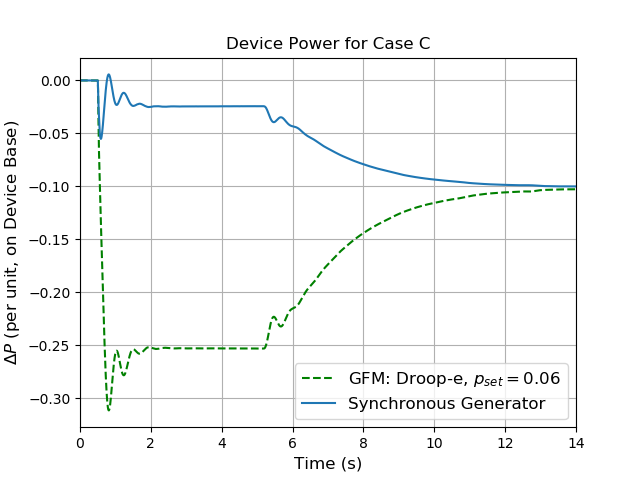}
		\caption{\small Case C}\label{fig: power c}
	\end{subfigure}
	\hfill
	\begin{subfigure}[t]{2.35in}
	    \centering
	    \includegraphics[trim=3 6 38 37,clip, width=1\textwidth]{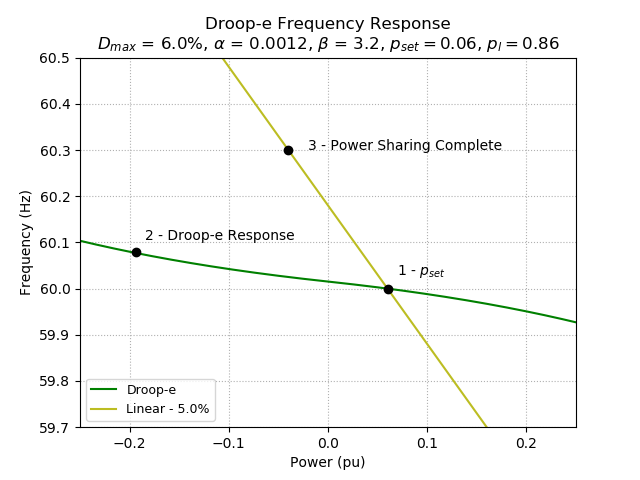}
	    \caption{\small Case C - GFM Droop Trajectory}\label{fig: droop c}
	\end{subfigure}
	\caption{The frequency response, power response, and droop trajectory for the GFM and SG devices in Case C.}\label{fig: Case C 3bus}
\end{figure*}

The results of case A are shown in Fig. \ref{fig: Case A 3bus}, with frequency statistics in Table \ref{tab: cases results}. The frequency plot shows a highly damped ($\zeta = 0.17$) response of the slow frequency mode ($\lambda_{5,6}$), with a ROCOF of 0.77 Hz/s and a nadir of 59.9 Hz. At approximately 5 seconds, the autonomous power sharing controller engages after sensing the disturbance and a return to steady state, which moves both devices to a 5\% equitable power sharing point. The results of a 5\% linear droop control are shown for comparison; the \textit{Droop-e} control yields a superior frequency response. The device powers (Fig. \ref{fig: power a}) show that the majority of the load perturbation was met by the GFM inverter, which provides a large effective inertia at the low dispatch of $p_{set} = 0.06$. Fig. \ref{fig: droop a} shows the position on the \textit{Droop-e} curve at $p_{set}$ (point 1), the change in frequency with power along the \textit{Droop-e} curve following the perturbation (point 2), and finally the position of the device on the 5\% power sharing curve after the autonomous controller action is complete.

Case B, with results in Fig. \ref{fig: Case B 3bus}, has a $p_{set}$ of $0.8$. There is a larger frequency deviation (Fig. \ref{fig: freq b}) as compared to case A, because the device is operating much closer to the power maximum. The ROCOF is larger, at 1.48 Hz/s, which is expected due to smaller effective inertia at the higher dispatch. The low frequency oscillatory element has decreased damping ($\zeta = 0.36$), as expected from the $\lambda_{5,6}$ power dispatch trajectory (Fig. \ref{fig: eigen 3}). The power outputs in Fig. \ref{fig: power b} show that the SG delivers a larger amount of power, only because the frequency has a larger deviation from the nominal. The movement on the \textit{Droop-e} curve in Fig. \ref{fig: droop b} shows that at this dispatch, the control moves past the linearization point ($p_l$), at which point a linear 6\% droop is applied. This is corroborated by the lower frequency as compared to the 5\% control simulation, and the decrease power delivery. However, this proves that the linearization control works and limits the output of the device.

The impact of a 20\% load decrease at a dispatch of $p_{set} = 0.06$ for case C is shown in Fig. \ref{fig: Case C 3bus}. This represents an over-frequency event, and highlights the stability of the \textit{Droop-e} controller in a transition from a positive to a negative power output. The frequency response in Fig. \ref{fig: freq c} shows a very small increase in frequency (0.09 Hz) with excellent damping (0.52), prior to the autonomous power sharing controller action. Because the \textit{Droop-e} curve is passing through the origin, which is the power output with the largest effective inertia, the smaller ROCOF of 0.68 as compared to case A is expected. The GFM sinks the majority of the surplus power (Fig. \ref{fig: power c}), while the movement on the \textit{Droop-e} curve shows the expected response (Fig. \ref{fig: droop c}), prior to achieving equitable power sharing with operation in over frequency domains. These three time-domain simulations show the successful operation of the \textit{Droop-e} controller at critical points, the linearization near the power limits and the inversion at the origin, as well as the superiority of this GFM control method.

\subsection{IEEE 39-Bus System}
\label{sec: numerical simulations}

Simulations on the IEEE 39-bus test system \cite{manitoba_hydro_international_ltd_pscad_nodate-1} demonstrate the capability of the \textit{Droop-e} control with multiple GFM devices. All devices are rated at 1000 MVA. The same dynamic SG model from the 3-bus simulations were used, with the rating changed. The dispatch of the devices is shown in Fig. \ref{tab:39 Bus config}. Note that Gen 0 is operating at full output. Three cases were developed; (i) \textit{Case 39-A} in which all generators were SGs, (ii) \textit{39-B} where three of the generators, 0, 4, and 8, were replaced with GFMs with a linear-5\% droop control, and  (iii) \textit{Case 39-C} where those three generators, 0, 4, and 8, were replaced with GFMs with \textit{Droop-e} control. In all three cases, the dispatch of the devices remained identical, which are summarized in Table \ref{tab:39 Bus config}. The perturbation applied is a disconnection of generator 7.

\begin{figure}[ht]
    \centering
    \includegraphics[width=0.95\columnwidth,trim={8 8 35 37},clip]{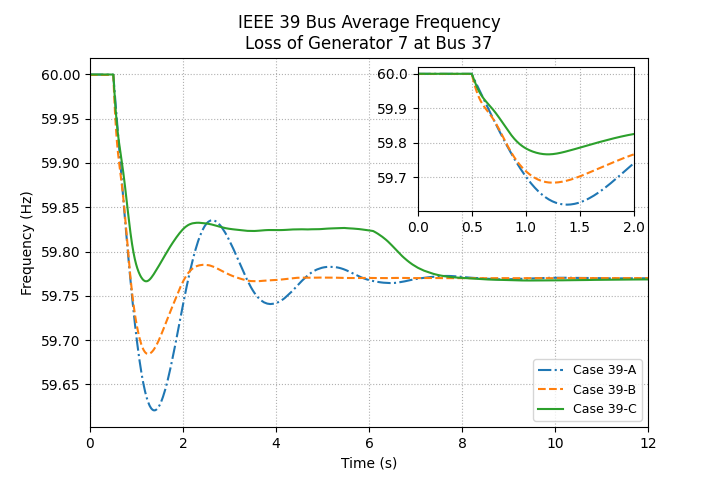}
    \caption{\small Average frequency response of IEEE 39-bus system for the three cases simulated following the loss of generator 7}
    \label{fig: 39 bus freq}
\end{figure}

Fig. \ref{fig: 39 bus freq} shows the average frequency response of the system for the loss of generator 7 for each of the three cases studied. Corresponding frequency metrics are presented in Table \ref{tab: 39 cases results}. The frequency response in Case 39-A is a classic second-order response of an SG dominated power system; the nadir (59.62 Hz) was substantially lower than the settling frequency (59.77 Hz), and an inter-area oscillatory mode at 0.44 Hz is present. In Case 39-B, the frequency response is improved relative to Case 39-A. The nadir is raised to 59.68 Hz, but the ROCOF is higher at 0.87 Hz/s as compared to 0.66 Hz/s for Case 39-A, which is expected according to \cite{kenyon_interactive_2023}. The frequency response with three \textit{Droop-e} GFM devices on the system in Case 39-C is the best of all three cases. The nadir (59.77 Hz) is equal to the settling frequency, which occurs because the \textit{Droop-e} equipped GFM devices deliver more power to the system during the initial transient, arresting the frequency decline faster than the linear droop control. However, the ROCOF is also the same as the SG-only case at 0.66 Hz/s, which occurs because of the improved effective inertia of the \textit{Droop-e} control at these dispatches. 

\begin{figure}[htb!]
    \centering
    \includegraphics[width=0.9\columnwidth,trim={10 10 10 35},clip]{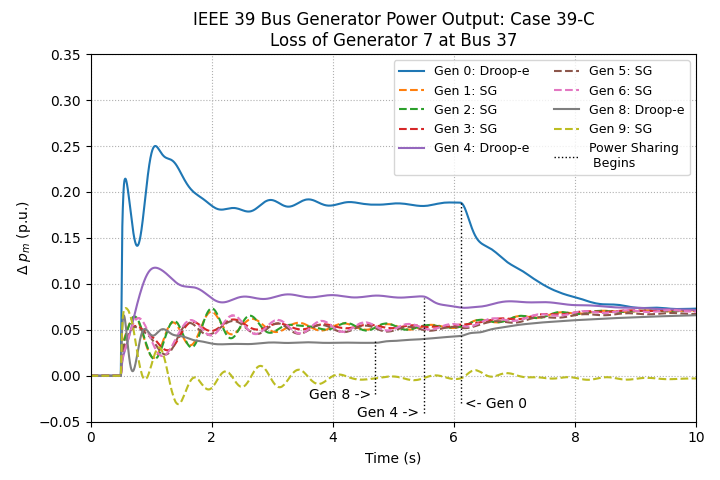}
    \caption{\small Per unit active power deviations following the loss of generator 7 on the 39-bus system, case 39-C; the initiation time for power sharing of each \textit{Droop-e} generator is highlighted.}
    \label{fig: 39 bus power delta}
\end{figure}

Fig. \ref{fig: 39 bus power delta} displays the power outputs of all of the generators in Case 39-C. Clearly, the \textit{Droop-e} controlled GFM devices release more power immediately following the disturbance (coinciding with when the additionally power is needed the most) and, as a result, provide more robust support to the grid during the transient period. This observation highlights the control benefit of leveraging the available headroom that is not deployed with a linear droop control; generator 0 delivered almost 3 times as much power to the network as a \textit{Droop-e} device compared to the other devices. Generator 4 delivers approximately 30\% more power, while generator 8 delivers less power, as the control objective for devices near the maximum power is to limit the output. 

\begin{table}[tb]
    \footnotesize
    \renewcommand{\arraystretch}{1.2}
    \setlength{\tabcolsep}{.3em}
    \centering
    \caption{\small Case Results--Frequency Statistics Derived From Average}
    \label{tab: 39 cases results}
    \begin{tabular}{c||c|c|c|c|c}
        \multirow{2}{*}{\rotatebox{0}{Case}} & Inertia & Nadir & ROCOF & Damping & Dom. Mode\\
        & (s) & (Hz) & (Hz/s) & ($\zeta$) & Freq. (Hz)\\
        \hline\hline
        39-A & 3.0 & 59.62 & 0.66 & 0.10 & 0.40\\
        39-B & 2.1 & 59.68 & 0.87 & 0.15 & 0.43\\
        39-C & 2.1 & 59.77 & 0.66 & 0.16 & 0.44\\
    \end{tabular}
\end{table}

The autonomous power sharing initiation times for \textit{Droop-e} controlled devices 0, 4, and 8, occurred at 6.2, 5.5, and 4.7 seconds, respectively, as seen by the indicators in Fig. \ref{fig: 39 bus power delta}.  Because the power sharing controller is activated by local measurements, the activation criteria is met at different times. The change in power output is a function of the local device deviation from the specified power sharing droop, and therefore a different power rate of change is expected for each \textit{Droop-e} device. The novelty of the power sharing controller is such that despite distributed generation devices with disparate activation times and a variety in delivered transient power, all devices arrive at the target power sharing droop value autonomously, with zero communication.

\section{Conclusion}
\label{sec:conclusion}

In future power systems dominated by renewable power generators, the presence of battery energy storage systems at a variety of dispatches will grow in prevalence. This paper presented the novel \textit{Droop-e} control strategy for grid-forming power electronic converters that establishes a non-linear active power--frequency relationship based on an exponential function of the power output. The advantages of this control approach consist of an increased utilization of available headroom to mitigate frequency excursions and the rate of change of frequency, accompanied by a natural limiting behavior. An auxiliary controller achieves autonomous power sharing following the initial transient stabilization. The motivation and design of the \textit{Droop-e} control was presented, and the stability of the controller was proven analytically, and with electromagnetic transient time-domain simulations. The simulations on a larger network established the autonomous, and superior, frequency stability of networks with \textit{Droop-e} power-electronic converters. Potential directions for future research are identified as:
\begin{itemize}
    \item Additional stability analyses, both analytical and large-signal, of larger networks with a more diverse set of frequency responsive devices.
    \item Optimality approaches for the tuning of the $\alpha$ and $\beta$ parameters.
    \item Investigation of \textit{Droop-e} devices at the interconnection level, where the larger headroom utility may achieve frequency response obligations with fewer devices.
\end{itemize}

\bibliographystyle{IEEEtran}
{\footnotesize\bibliography{references}}
\section{Appendix}
\label{sec:appendix}

Tables \ref{tab: system parameters} and \ref{tab:39 Bus config} provide the detailed device and system parameters, and the dispatch of the 39-bus simulation cases, respectively.

\begin{table}[htbp]
    \footnotesize
    \renewcommand{\arraystretch}{1.0}
    \setlength{\tabcolsep}{.1em}
    \centering
    \caption{\small Device and Network Parameters}
    \begin{tabular}{c|c||c|c||c|c}
    \textbf{Parameter} & \textbf{Value} & \textbf{Par...} & \textbf{Value} & \textbf{Par...} & \textbf{Value}\\\hline\hline
    $H(secs)$ & 3.01 & $X_d(pu)$ & 1.3125 & $X'_d(pu)$ & 0.1813\\\hline
    $X_q$ & 1.2578 & $X'_q(pu)$ & 0.25 & $T'_{do}(sec)$ & 5.89\\\hline
    $T'_{qo}(sec)$ & 0.6 & $K_A$ & 20 & $T_A(sec)$ & 0.2\\\hline
    $K_E$ & 1.0 & $T_E(sec)$ & 0.314 & $K_F$ & 0.063\\\hline
    $T_F(sec)$ & 0.35 & $S_E-\gamma$ & 0.0039 & $S_E-\epsilon$ & 1.555\\\hline
    $M_D(\%)$ & 5 & $\omega_B(rad/s)$ & 377 & $X_a(pu)$ & 0.05\\\hline
    $X_b(pu)$ & 0.05 & $D_I-\alpha$ & 0.002 & $D_I-\beta$ & 3.0\\\hline
    $X_{GFM}(pu)$ & 0.15 & $R_{GFM}(pu)$ & 0.005 & $T_{fil}(sec)$ & 0.0167\\\hline
    $P_2(pu)$ & 0.75 & $Q_2(pu)$ & 0.25 & $V_1(pu)$ & 1.02\\\hline
    $V_3(pu)$ & 1.02 & $S_{G}(MVA)$ & 100 & $S_{I}(MVA)$ & 50\\\hline
    $L_f(pu)$ & 0.15 & $R_f(pu)$ & 0.005 & $C_f(pu)$ & 2.5\\\hline
    $R_{cap}(pu)$ & 0.005 & $k_C^i$ & 1.19 & $k_C^p$ & 0.73\\\hline
    $G_C$ & 1.0 & $k_V^i$ & 1.16 & $k_V^p$ & 0.52 \\\hline
    $G_V$ & 1.0 & \phantom{0} & \phantom{0} & \phantom{0} & \phantom{0}\\
    \end{tabular}
    \label{tab: system parameters}
\end{table}

\begin{table}[htbp]
    \footnotesize
    \centering
    \caption{\small IEEE 39-Bus Configuration and Results}
    \setlength\tabcolsep{3.5pt}   
    \begin{tabular}{c|c|c|c|c}
    \multirow{2}{*}{\rotatebox{0}{Generator}} & P & \multicolumn{3}{c}{Generator Type for Case}\\
     & (MW) & 39-A & 39-B & 39-C \\\hline\hline
    0 & 250 & SG & Linear-5\% & \textit{Droop-e}\\
    1 & 678 & SG & SG & SG\\
    2 & 650 & SG & SG & SG\\
    3 & 632 & SG & SG & SG\\
    4 & 508 & SG & Linear-5\% & \textit{Droop-e}\\
    5 & 650 & SG & SG & SG\\
    6 & 560 & SG & SG & SG\\
    7 & 540 & SG & SG & SG\\
    8 & 830 & SG & Linear-5\% & \textit{Droop-e}\\
    9 & 1000 & SG & SG & SG\\
    \end{tabular}
    \label{tab:39 Bus config}
\end{table}

% \begin{table}[htbp]
%     \small
%     \renewcommand{\arraystretch}{1.2}
%     \setlength{\tabcolsep}{.4em}
%     \centering
%     \caption{Additional Dynamic System Parameters}
%     \begin{tabular}{c|c||c|c||c|c}
%     \textbf{Parameter} & \textbf{Value} & \textbf{Par...} & \textbf{Value} & \textbf{Par...} & \textbf{Value}\\\hline\hline
%     $L_f(pu)$ & 0.15 & $R_f(pu)$ & 0.005 & $C_f(pu)$ & 2.5\\\hline
%     $R_{cap}(pu)$ & 0.005 & $k_C^i$ & 1.19 & $k_C^p$ & 0.73\\\hline
%     $G_C$ & 1.0 & $k_V^i$ & 1.16 & $k_V^p$ & 0.52 \\\hline
%     $G_V$ & 1.0 & \phantom{0} & \phantom{0} & \phantom{0} & \phantom{0}\\
%     \end{tabular}
%     \label{tab: PSCAD system parameters}
% \end{table}

% that's all folks
\end{document}